\documentclass{article}
\title{Singlet states and the estimation of eigenstates and 
eigenvalues of an unknown Controlled-U gate}
\author{Mark Hillery \\ Department of Physics \\ Hunter College of
CUNY \\ 695 Park Avenue \\ New York, NY 10021 \and
Vladim\'{i}r Bu\v{z}ek \\ Institute of Physics, Slovak Academy of
Sciences \\ D\'{u}bravsk\'{a} cesta 9 \\ 842 28 Bratislava,
Slovakia \\ and \\ Faculty of Informatics, Masaryk University \\
Botanick\'{a} 68a \\ 602 00 Brno, Czech Republic} 
\begin{document}
\maketitle
\begin{abstract}
We consider several problems that involve finding the eigenvalues
and generating the eigenstates of unknown unitary gates.  We first
examine Controlled-U gates that act on qubits and assume that we
know the eigenvalues.  It is then shown how to use singlet states
to produce qubits in the eigenstates of the gate.  We then remove
the assumption that we know the eigenvalues and show how to both
find the eigenvalues and produce qubits in the eigenstates.  
Finally, we look at the case where the unitary operation acts on
qutrits and it has two eigenvalues of $1$ and one of $-1$.  We
are able to use a singlet state to produce a qutrit in the 
eigenstate corresponding to the $-1$ eigenvalue.
\end{abstract}
\section{Introduction}
A common problem, which arises in quantum mechanics, is finding
the eigenvalues and eigenstates of an operator, usually the
Hamiltonian.  The eigenvalues are the values
that the observable corresponding to the operator can assume, and
the eigenstates are the states of the system in which that
observable will have a definite value.

With the advent of quantum algorithms, a natural question to raise
is whether there are quantum algorithms that will efficiently find 
the eigenvalues and eigenvectors of operators.  The answer to this
question is, in fact, yes.  Based on earlier work by Kitaev \cite
{kitaev}, Cleve, et al.\ developed an algorithm that can estimate
an eigenvalue if one copy of the eigenstate is provided \cite{cleve}.
This algorithm was analyzed further by Abrams and Lloyd 
\cite{abrams}.  They pointed out that it is not necessary to have
a copy of an eigenstate to use this procedure.  One can start with
an arbitrary input state, and at the end of the procedure, one will 
obtain an eigenvalue corresponding to an eigenstate that has a
nonzero overlap with the input state.  This is not a deterministic
procedure; we could obtain any eigenvalue whose eigenstate has a
significant overlap with the input vector.  In addition, Abrams
and Lloyd showed that at the
output one has not only the eigenvalue, but a set of qubits that
is in a state that is a good approximation to the eigenstate corresponding to the measured eigenvalue.  How good this 
approximation is was recently investigated by Travaglione 
and Milburn \cite{milburn}. 

This procedure requires that one have some knowledge about the
eigenstate one is trying to generate and whose eigenvalue one
is trying to find.  In particular, it is necessary that
the input state have a substantial overlap with the desired
eigenstate.  It may be possible to accomplish this if there is 
some information available that allows a guess for the state to 
be made.  For example, when finding the ground state energy of
a not-too-complicated Hamiltonian, it might be possible on 
physical grounds to obtain a rough idea of what the ground
state would look like, and this information could be used to
design an appropriate input state for the eigenvalue
estimation algorithm. In many cases, however, there will be
little if any information to guide ones choice, with the result
that the input state may have a very small or no overlap with
the desired eigenstate.

Here we shall show that a different input state, a singlet state,
allows one to find all eigenvalues and eigenstates of an
unknown Controlled-U gate simultaneously.  We shall start with
the case of a single qubit gate where we know the eigenvalues
and wish to generate output qubits in the eigenstates of the
gate.  We shall then proceed to the case where the gate still
operates on only a single qubit, but we do not know either its
eigenvalues or its eigenstates.  Our object then is to find the
eigenvalues and produce output qubits in the eigenstates.  Finally,
we shall consider a Controlled-U gate that acts on qutrits and has
two eigenvalues of $1$ and one eigenvalue of $-1$.  It is possible
to use a singlet state to produce a qutrit in the eigenstate
corresponding to the eigenvalue $-1$ with a network that contains
only two Controlled-U gates.  This procedure is easily generalized 
to $D$-dimensional quantum systems, qudits.  Given a Controlled-U
gate that acts on qudits and has eigenvalues $1$, which is 
$(D-1)$-fold degenerate, and $-1$, it is possible to produce a
qudit in the eigenstate with eigenvalue $-1$ by using a network
containing $D-1$ Controlled-U gates.

\section{Generation of eigenstates of a Controlled-$U$ gate with
known eigenvalues}
Consider the following problem.  We are given a
Controlled-$U$ gate which acts on single qubits and we would 
like to generate its eigenstates.  We know that the 
eigenvalues of the gate are $1$ and $-1$, but have no
information about its eigenstates.  A measurement-based
strategy for doing this would involve sending qubits
through the gate and measuring them.  For example, we
could use the basis states $|0\rangle$ and $|1\rangle$
to obtain information about the matrix elements of $U$
in that basis.  If we send the state
$|1\rangle_{a}|0\rangle_{b}$, where $a$ is the
control bit and $b$ is the target bit,  
through the gate, the probability that target qubit
at the output is in the state $|0\rangle_{b}$
can be measured.  This probability is just equal to
$|\langle 0|U|0\rangle |^{2}$.  The probability that
the target qubit is in the state $|1\rangle_{b}$ is
just $|\langle 1|U|0\rangle |^{2}$.  These measurements
give us information about two of the matrix elements
of $U$, and the fact that we know that the eigenvalues
are $1$ and $-1$ means that these are the only two we have
to know.  In particular we have that
\begin{equation}
\langle 0|U|0\rangle = -\langle 1|U|1\rangle \hspace{1cm}
\langle 0|U|1\rangle = \langle 1|U|0\rangle^{\ast} .
\end{equation}
Information about the relative phase of these matrix
elements can be gained by using the input state
$(|0\rangle +e^{i\theta}|1\rangle )/\sqrt{2}$.  The
probability, $p_{0}$ that the output vector is in the
state $|0\rangle$ is 
\begin{equation}
p_{0}=|\langle 0|U|0\rangle +e^{i\theta}\langle 0|U|
1\rangle |^{2}.
\end{equation}
After sending through many qubits we would have an estimate
of the matrix elements, and we could then diagonalize the
matrix.  This information could then be used to generate
qubits in the eigenstates of $U$.  This procedure
involves many qubits and many uses of the gate.  What we 
shall now present
is a quantum strategy that will produce both
eigenstates with certainty using only 3 qubits and requiring
only one use of the gate.

Let $a$ and $b$ be the control and target qubits of the
gate, as before.  To this we add a third qubit, which we
shall denote by $c$.  We now define the following states
\begin{eqnarray}
|\pm x\rangle &=& \frac{1}{\sqrt{2}}(|0\rangle \pm 
|1\rangle ) \nonumber \\
|\phi_{s}\rangle &=& \frac{1}{\sqrt{2}}(|01\rangle
-|10\rangle ) ,
\end{eqnarray}
and note that the rotational invariance of the singlet
state, $|\phi_{s}\rangle$, implies that it can also
be expressed as
\begin{equation}
\label{sing1}
|\phi_{s}\rangle = \frac{1}{\sqrt{2}}(|u_{+}\rangle
|u_{-}\rangle - |u_{-}\rangle |u_{+}\rangle ) ,
\end{equation}
where $|u_{+}\rangle$ is the eigenstate of $U$ with 
eigenvalue $1$ and $|u_{-}\rangle$ is the eigenstate with
eigenvalue $-1$.  More explicitly, the transformation
specified by
\begin{equation}
V|0\rangle = |u_{+}\rangle \hspace{1cm} V|1\rangle = 
|u_{-}\rangle ,
\end{equation}
is unitary, and, because $|\phi_{s}\rangle$ is invariant
under $U(2)\otimes U(2)$, we have
\begin{equation}
\label{sing2}
(V\otimes V)|\phi_{s}\rangle = |\phi_{s}\rangle .
\end{equation}
Because the phase of the eigenstates is
arbitrary, in general Eqs.\ (\ref{sing1}) and (\ref{sing2})
will be true only up to an overall phase factor, but we shall 
assume that the phases of $|u_{+}\rangle$ and $|u_{-}\rangle$ 
are such that the phase factor is equal to one.

We now start our $3$-qubit system in the state
\begin{equation}
|\Psi_{in}\rangle_{abc}=|+x\rangle_{a}|\phi_{s}
\rangle_{bc} .
\end{equation}
After qubits $a$ and $b$ go through the Controlled-$U$
gate the state of the system is
\begin{eqnarray}
|\Psi_{out}\rangle_{abc} &=& \frac{1}{2}[|0\rangle_{a}
(|u_{+}\rangle_{b}|u_{-}\rangle_{c} 
- |u_{-}\rangle_{b} |u_{+}\rangle_{c}) \nonumber \\
 & & +|1\rangle_{a}(|u_{+}\rangle_{b}|u_{-}\rangle_{c} +|u_{-}\rangle_{b} |u_{+}\rangle_{c})] .
\end{eqnarray}
We now want to measure the $a$ qubit in the $|\pm x\rangle$
basis.  In order to see the result of doing so we can
express $|\Psi_{out}\rangle_{abc}$ as
\begin{equation}
|\Psi_{out}\rangle_{abc}=\frac{1}{\sqrt{2}} (|+x\rangle_{a}
|u_{+}\rangle_{b}|u_{-}\rangle_{c}-|-x\rangle_{a}
|u_{-}\rangle_{b}|u_{+}\rangle_{c}) .
\end{equation}
This equation implies that if we measure $a$ and get 
$|+x\rangle$, then qubit $b$ is in the $1$ eigenstate and
qubit $c$ is in the $-1$ eigenstate, while if we get
$|-x\rangle$, then it is just the other way around.
Therefore, with one use of the gate we have, with certainty,
generated its two eigenstates.

Now let us see what happens when the eigenvalues are not 
$\pm 1$ but two other complex numbers of modulus unity.
We shall denote the eigenstates as $|u_{1}\rangle$ and 
$|u_{2}\rangle$ and the corresponding eigenvalues as 
as $e^{i\theta_{1}}$ and $e^{i\theta_{1}}$, respectively.  
We again use the same $3$-qubit scheme and choose the input 
state to be $|+x\rangle_{a}|\phi_{s}\rangle_{bc}$.  The 
output state is now given by
\begin{equation}
|\Psi_{out}\rangle_{abc}=\frac{1}{\sqrt{2}}(|v_{1}\rangle_{a}
|u_{1}\rangle_{b}|u_{2}\rangle_{c}-|v_{2}\rangle_{a}
|u_{2}\rangle_{b}|u_{1}\rangle_{c}) ,
\end{equation}
where
\begin{eqnarray}
|v_{1}\rangle &=& \frac{1}{\sqrt{2}}(|0\rangle +e^{i\theta_{1}}
|1\rangle ) \nonumber \\
|v_{2}\rangle &=& \frac{1}{\sqrt{2}}(|0\rangle +e^{i\theta_{2}}
|1\rangle ).
\end{eqnarray}
Note that $|v_{1}\rangle$ and $|v_{2}\rangle$ are not orthogonal,
but because we assume we know the eigenvalues, these vectors
are known.  At this point we can apply the optimal procedure for distinguishing two nonorthogonal states \cite{ivanovic}-\cite{peres}.  
This is a generalized measurement that can be applied when we are
given a state, which is one of two known states, and we want to
determine which of the two states it is.  The measurement has one
of three possible outcomes; it either tells us without error which 
of the two states we have, or it tells us that it has failed to 
distinguish the states.  Applied to $|v_{1}\rangle$ and 
$|v_{2}\rangle$, the procedure would succeed with a probability
of
\begin{equation}
1-|\langle v_{1}|v_{2}\rangle |=1-\frac{1}{\sqrt{2}}[1+\cos
(\theta_{1}-\theta_{2})]^{1/2} .
\end{equation}
The closer the phase difference between the eigenvalues is to
$\pi$, the greater the probability of success of this procedure.
If $|v_{1}\rangle$ is detected at output $a$, then $|u_{1}\rangle$
is at output $b$ and $|u_{2}\rangle$ at $c$.  On the other hand, if 
$|v_{2}\rangle$ is detected at $a$, then $|u_{2}\rangle$ is
at $b$ and $|u_{1}\rangle$ is at $c$.

If one allows more than one use of the gate there are other
possibilities.  Suppose that we know the eigenvalues of $U$ are
$1$ and $i$.  Then the eigenvalues of $U^{2}$ are $1$ and $-1$.
We can then apply the above procedure to generate the eigenstates
with one modification.  The three-qubit initial state is the 
same, but qubits $a$ and $b$ pass through two Controlled-$U$
gates instead of one.  Again qubit $a$ is measured at the
output in the $|\pm x\rangle$ basis.  If $|+x\rangle$ is 
found, then the eigenstate corresponding to $1$ is at output
$b$ and that corresponding to $i$ is at output $c$.  If
$|-x\rangle$ is found, the $b$ and $c$ outputs are reversed.

As preparation for the next section, let us consider one last,
harder problem.  Suppose that we know that the eigenvalues are
not the same, and that they are members of the set $\{ 1,-1,
i,-i\}$.  In this case we have partial rather than complete
information about the eigenvalues, and we again want to 
generate qubits in the eigenstates.  This can be done using
four qubits, a Controlled-$U$ gate, and a Controlled-$U^{2}$
gate. This last gate can be constructed from two Controlled-$U$
gates in sequence.  Qubit $a$ is the control bit for the
Controlled-$U^{2}$ gate, qubit $b$ is the control bit for the 
Controlled-$U$ gate, and qubit $c$ is the target bit for both.
Let $|u_{1}\rangle$ and $|u_{2}\rangle$ be the eigenstates
of $U$ with corresponding eigenvalues $z_{1}$ and $z_{2}$,
where $z_{1}$ and $z_{2}$ are members of the set $\{1,-1,
i,-i\}$.  The input state is
\begin{eqnarray}
|\Psi_{in}\rangle_{abcd} & = &|+x\rangle_{a} |+x\rangle_{b}
|\phi_{s}\rangle_{cd} \nonumber \\
 & = & |+x\rangle_{a} |+x\rangle_{b}\left(\frac{1}
{\sqrt{2}}\right) (|u_{1}\rangle_{c}|u_{2}\rangle_{d}
-|u_{2}\rangle_{c}|u_{1}\rangle_{d}) .
\end{eqnarray}
The output state of the network is given by
\begin{eqnarray}
|\Psi_{out}\rangle_{abcd} & = &G_{ac}(U^{2})G_{bc}(U)
|\Psi_{in}\rangle_{abcd} \nonumber \\
 & = &\frac{1}{2\sqrt{2}}[(|00\rangle_{ab}+z_{1}|01\rangle_{ab}
+z_{1}^{2}|10\rangle_{ab}+z_{1}^{3}|11\rangle_{ab})
|u_{1}\rangle_{c}|u_{2}\rangle_{d} \nonumber \\
 & & -(|00\rangle_{ab}+z_{2}|01\rangle_{ab}
+z_{2}^{2}|10\rangle_{ab}+z_{2}^{3}|11\rangle_{ab})
|u_{2}\rangle_{c}|u_{1}\rangle_{d}] ,
\end{eqnarray}
where $G_{jk}(U^{n})$ is the operator corresponding to a
Controlled-$U^{n}$ gate with control bit $j$ and target
bit $k$. Now consider the vector
\begin{equation}
|\eta (z)\rangle_{ab}=\frac{1}{2}(|00\rangle_{ab}+z
|01\rangle_{ab}+z^{2}|10\rangle_{ab}+z^{3}|11\rangle_{ab}) .
\end{equation}
The vectors $|\eta (1)\rangle_{ab}$, $|\eta (-1)\rangle_{ab}$, 
$|\eta (i)\rangle_{ab}$, and $|\eta (-i)\rangle_{ab}$ form
an orthonormal basis of the space of the two qubits $a$ and
$b$.  If we now measure the two qubit system $ab$ in this 
basis, we can determine one of the eigenvalues of $U$, e.g.\ 
if the result of the measurement is $|\eta (1)\rangle_{ab}$,
then one of the eigenvalues is $1$.  The eigenstate 
corresponding to this eigenvalue will emerge from output $c$,
and the eigenstate corresponding to the other, unknown,
eigenvalue will emerge from output $d$.

This procedure will allow us to find one of the eigenvalues, if
we know that they belong to a limited set, and generate both
eigenstates.  A better procedure would allow us to find both
eigenvalues, remove the restriction that they belong to a
particular set, and generate both eigenstates.  Such an
algorithm is presented in the next section. 

\section{Application of phase estimation to find unknown
eigenvalues and eigenvectors}
Suppose that we have an unkown Controlled-U gate, and we
want to find its eigenvalues and generate qubits in its 
eigenstates.  This can be done by modifying the phase estimation
algorithm of Cleve, et al.\ and using a singlet state as the 
input \cite{cleve}.  One takes two phase-estimation circuits
for the same gate, and sends into each one of two particles,
which, together, form a singlet state.  This avoids the main disadvantage of the original algorithm.  There, besides the
Controlled-U gates, one also needed a qubit prepared in one
of the eigenstates.  Sending this qubit through the network
would then generate an estimate of the eigenvalue for this
eigenstate.  An alternative is to send in a random qubit,
in which case one gets an estimate for a random eigenvalue. 
In particular, the estimate corresponds to one of the eigenvalues
whose eigenstates have a nonzero overlap with the input state.
The original qubit is left in a state that is close approximation
to the eigenstate corresponding to the measured eigenvalue 
\cite{abrams,milburn}.

Each of the two networks, which we shall label $A$ and $B$ is constructed as follows.  We have $n$ control qubits, which
for network $A$ we shall call $A1, A2, \ldots An$, and one target
bit, which we shall call $A$.  Each of the control bits is initially
in the state $(|0\rangle +|1\rangle )/\sqrt{2}$.  Control bit $Aj$
is connected to a gate that does nothing if the control bit is $0$,
and performs the operation $U^{2^{j-1}}$ if the control bit is $1$.
The network $B$ is identical.  The effect of the entire network is
given by
\begin{equation}
|\Psi_{out}\rangle = G_{(Bn)B}(U^{2^{n-1}})\ldots G_{(B1)B}(U)
G_{(An)A}(U^{2^{n-1}})\ldots G_{(A1)A}(U)|\Psi_{in}\rangle .
\end{equation}
Let the eigenstates of $U$ be
$|u_{1}\rangle$ and $|u_{2}\rangle$ with eigenvalues $e^{i\phi_{1}}$ 
and $e^{i\phi_{2}}$, respectively.  As before, the singlet state can 
be expressed in terms of these eigenstates 
\begin{equation}
|\phi_{s}\rangle_{AB} = \frac{1}{\sqrt{2}}(|u_{1}\rangle_{A}
|u_{2}\rangle_{B}-|u_{2}\rangle_{A}|u_{1}\rangle_{B}) .
\end{equation}
The initial state of the system is then
\begin{equation}
|\Psi_{in}\rangle = \frac{1}{\sqrt{2}2^{n}}( 
|u_{1}\rangle_{A}|u_{2}\rangle_{B}-|u_{2}\rangle_{A}|u_{1}
\rangle_{B})
\prod_{j=0}^{n-1}(|0\rangle_{Aj}+|1\rangle_{Aj})
\prod_{k=0}^{n-1}(|0\rangle_{Bk}+|1\rangle_{Bk}) .
\end{equation}
After passing through the networks this state becomes
\begin{eqnarray}
|\Psi_{in}\rangle & \rightarrow & \frac{1}{\sqrt{2}2^{n}}
\left[ |u_{1}\rangle_{A}\prod_{j=0}^{n-1}(|0\rangle_{Aj}+
e^{i2^{j}\phi_{1}}|1\rangle_{Aj})|u_{2}\rangle_{B}\prod_{k=0}^{n-1}
(|0\rangle_{Bk}+e^{i2^{k}\phi_{2}}|1\rangle_{Bk}) \right.
\nonumber \\
 & & \left. -|u_{2}\rangle_{A}\prod_{j=0}^{n-1}(|0\rangle_{Aj}+
e^{i2^{j}\phi_{2}}|1\rangle_{Aj})|u_{1}\rangle_{B}\prod_{k=0}^{n-1}
(|0\rangle_{Bk}+e^{i2^{k}\phi_{1}}|1\rangle_{Bk})\right] .
\end{eqnarray}
The products in the above equation can be expressed as a sum over
$n$-digit binary numbers.  For example,
\begin{equation}
\prod_{j=0}^{n-1}(|0\rangle_{Aj}+e^{i2^{j}\phi_{1}}|1\rangle_{Aj})
=\sum_{y=0}^{2^{n}-1}e^{i\phi_{1}y}|y\rangle_{An,\ldots A1} .
\end{equation}
The first digit of the $n$-digit binary number $y$ corresponds to 
the state of system $An$, the second to that of $A(n-1)$, and so
on.  In the above equation we have indicated this explicitly with
subscribts on the state, but if the future these will be omitted
and this correspondence will be understood.  It is still necessary
to indiciate whether $|y\rangle$ is a state of $An, \ldots A1$ or
$Bn, \ldots B1$, and this will be indicated by the subscripts 
$\overline{A}$ and $\overline{B}$, respectively.  We then have that
\begin{eqnarray}
|\Psi_{initial}\rangle & \rightarrow & \frac{1}{\sqrt{2}2^{n}}
\left[ |u_{1}\rangle_{A}(\sum_{y=0}^{2^{n}-1}e^{i\phi_{1}y}
|y\rangle_{\overline{A}})|u_{2}\rangle_{B}(\sum_{w=0}^{2^{n}-1}
e^{i\phi_{2}w}|w\rangle_{\overline{B}}) \right. \nonumber \\
 & &\left. -|u_{2}\rangle_{A}(\sum_{y=0}^{2^{n}-1}e^{i\phi_{2}y}
|y\rangle_{\overline{A}})|u_{1}\rangle_{B}(\sum_{w=0}^{2^{n}-1}
e^{i\phi_{1}w}|w\rangle_{\overline{B}}) \right] .
\end{eqnarray}

The next step is to apply the quantum inverse Fourier transform
operation to states $\overline{A}$ and $\overline{B}$.  This takes
the state $|y\rangle$ to
\begin{equation}
|y\rangle \rightarrow \frac{1}{2^{n/2}}\sum_{z=0}^{2^{n}-1}
e^{-2\pi iyz/2^{n}}|z\rangle
\end{equation}
Before applying this, however, we want to express the phases
$\phi_{1}$ and $\phi_{2}$ in different way.  First, let
$x_{j}=\phi_{j}/(2\pi )$ for $j=1,2$, which implies that $0\le
x_{j}<1$.  In addition, let $\overline{x}_{j}$ be the closest
integer to $2^{n}x_{j}$ (we assume $\overline{x}_{j}$ is expressed 
in binary form) so that
\begin{equation}
x_{j}=\frac{\overline{x}_{j}}{2^{n}}+\delta_{j} ,
\end{equation}
where $|\delta_{j}|\le 1/2^{n+1}$.  If we now apply the inverse
Fourier transform we find that
\begin{equation}
\sum_{y=0}^{2^{n}-1}e^{2\pi iy[(\overline{x}_{j}/2^{n})+\delta_{j}]}
|y\rangle \rightarrow \frac{1}{2^{n/2}}\sum_{y=0}^{2^{n}-1}
\sum_{z=0}^{2^{n}-1}e^{2\pi iy(\overline{x}_{j}-z)/2^{n}}
e^{2\pi iy\delta_{j}}|z\rangle .
\end{equation}
It is possible to perform the $y$ summation in the above equation
\begin{eqnarray}
g(z;\overline{x}_{j},\delta_{j}) & = & \frac{1}{2^{n}}
\sum_{y=0}^{2^{n}-1}e^{2\pi iy(\overline{x}_{j}-z)/2^{n}}
e^{2\pi iy\delta_{j}} \nonumber \\
 & = &\frac{1}{2^{n}}\frac{1-e^{2\pi i\delta_{j}2^{n}}}{1-e^{2\pi 
i[((\overline{x}_{j}-z)/2^{n})+\delta_{j}]}} .
\end{eqnarray}
This function is peaked about $z=\overline{x}_{j}$ and the maximum
value of its magnitude
is greater than $2/\pi$ \cite{cleve}.  After applying
the inverse Fourier transform to both systems $\overline{A}$
and $\overline{B}$, our state is
\begin{eqnarray}
|\Psi_{out}\rangle & = & \frac{1}{\sqrt{2}} \left[ |u_{1}\rangle_{A}(\sum_{z=0}^{2^{n}-1}g(z;\overline{x}_{1},\delta_{1})
|z\rangle_{\overline{A}})|u_{2}\rangle_{B}(\sum_{s=0}^{2^{n}-1}
g(s;\overline{x}_{2},\delta_{2})|s\rangle_{\overline{B}}) \right. \nonumber \\
 & &\left. -|u_{2}\rangle_{A}(\sum_{z=0}^{2^{n}-1}
g(z;\overline{x}_{2},\delta_{2})|z\rangle_{\overline{A}})
|u_{1}\rangle_{B}(\sum_{s=0}^{2^{n}-1}g(s;\overline{x}_{1},
\delta_{1})|s\rangle_{\overline{B}}) \right] .
\end{eqnarray}

We now measure both systems $\overline{A}$ and $\overline{B}$
in the computational basis.  The most likely results are either
$\overline{x}_{1}$ for $\overline{A}$ and $\overline{x}_{2}$
for $\overline{B}$, in which case qubit $A$ is in $|u_{1}\rangle$
and $B$ is in $|u_{2}\rangle$, or $\overline{x}_{2}$ for 
$\overline{A}$ and $\overline{x}_{1}$ for $\overline{B}$, in which 
case qubit $A$ is in $|u_{2}\rangle$ and $B$ is in $|u_{1}\rangle$.
In either case we have both the eigenvalues (to n places in base 2)
and qubits in the eigenvectors.
  
\section{Higher dimensional systems}
The reasoning in the previous sections can be extended from qubits
to qudits, D-dimensional quantum systems.  The fully antisymmetric
state of D D-dimensional quantum systems is a $U(D)$ singlet
\cite{georgi}.  If we denote the computational basis states by $|n\rangle$, where $n=0,1,\ldots D-1$, this state can be 
expressed as
\begin{equation}
|\phi_{s}(D)\rangle = \frac{1}{\sqrt{D!}}
\sum_{j_{1}=0}^{D-1}\ldots\sum_{j_{D}=0}^{D-1}
\varepsilon_{j_{1}\ldots j_{D}}|j_{1}\rangle\ldots |j_{D}\rangle ,
\end{equation}
where $\varepsilon_{j_{1}\ldots j_{D}}$ is the totally 
antisymmetric tensor of rank $D$.  Now consider a unitary operator
$U$ whose eigenstates are $|u_{j}\rangle$, where $j=1,\ldots D$.
The fact that $|\phi_{s}(D)\rangle$ is a singlet means that it can 
be expressed as
\begin{equation}
|\phi_{s}(D)\rangle =\frac{1}{\sqrt{D!}}
e^{i\mu}\sum_{j_{1}=1}^{D}\ldots
\sum_{j_{D}=1}^{D}\varepsilon_{j_{1}\ldots j_{D}}
|u_{j_{1}}\rangle\ldots |u_{j_{D}}\rangle ,
\end{equation}
where $e^{i\mu}$ is a phase factor that depends on how the phases
of the eigenstates are chosen.  We shall subsequently assume that 
they have been chosen so that $\mu = 0$.

Let us now consider the following problem for the case $D=3$;
its generalization to the case of arbitrary dimension is
straightforward.  We are given a Controlled-$U$ gate, where 
the control is a qubit and the target is a qutrit. If the
control qubit is in the state $|0\rangle$ nothing happens
to the target qutrit, and if it is in the state $|1\rangle$,
the operation $U$ is performed on the qutrit. This
gate corresponds to the operator $G_{jk}(U)$, where $j$ is
the control qubit and $k$ is the target qutrit.  We shall
assume that the operator $U$ has eigenvalues $1$ and $-1$,
where the eigenvalue $1$ is degenerate, and we would like
to produce a qutrit in the eigenstate corresponding to $-1$.

This can be done with two Controlled-$U$ gates, two qubits,
and three qutrits.  The initial state of the system is
\begin{equation}
|\Psi_{in}\rangle_{a\dots e}=|+x\rangle_{a}|+x\rangle_{b}
|\phi_{s}(3)\rangle_{cde} .
\end{equation}
Particles $a$ and $b$ are qubits and $c$, $d$, and $e$ are
qutrits.  Qubit $a$ is the control bit for qutrit $c$ and qubit 
$b$ is the control bit for qutrit $d$.  The output state is
given by
\begin{equation}
|\Psi_{out}\rangle_{a\ldots e}=G_{ac}(U)G_{bd}(U)
|\Psi_{in}\rangle_{a\ldots e} .
\end{equation}
Let $|u_{1}\rangle$ and $|u_{2}\rangle$ be orthonormal
eigenstates of $U$ with eigenvalue $1$ and $|v\rangle$
be the eigenstate with eigenvalue $-1$.  In terms of these
states we have that
\begin{eqnarray}
|\Psi_{out}\rangle_{a\ldots e} & = & \frac{1}{\sqrt{6}}[
|-x\rangle_{a}|+x\rangle_{b}
(|vu_{1}u_{2}\rangle_{cde}-|vu_{2}u_{1}\rangle_{cde}) \nonumber \\
 & & +|+x\rangle_{a}|-x\rangle_{b}(|u_{2}vu_{1}\rangle_{cde}
-|u_{1}vu_{2}\rangle_{cde}) \nonumber \\
 & & +|+x\rangle_{a}|+x\rangle_{b}(|u_{1}u_{2}v\rangle_{cde}
-|u_{2}u_{1}v\rangle_{cde}) ].
\end{eqnarray}
We now measure qubits $a$ and $b$ in the $|\pm x\rangle$ basis.
If we find $a$ in the $|-x\rangle$ state, then qutrit $c$ is in
the eigenstate with eigenvalue $-1$, while if qubit $b$ is in
the $|-x\rangle$ state, then it is qutrit $d$ that is in the $-1$
eigenstate.  Finally, if both of these qubits is found to be in
the $|+x\rangle$ state, then qutrit $e$ is in the $-1$ eigenstate.

In the corresponding problem for qudits, $U$ has eigenvalues
$-1$, which is nondegenerate, and $1$, which is $D-1$ fold
degenerate.  The object is to produce a qudit in the eigenstate
corresponding to $-1$.  To do so one uses a network consisting
of $D-1$ qubits, $D$ qudits in a singlet state, and $D-1$ 
Controlled-$U$ gates.  The procedure is a simple generalization
of the one just discussed for qutrits.

\section{Conclusion}
We have shown that singlet states in combination with
Controlled-$U$ gates can be used to produce qubit, or qudits,
in eigenstates of the operator $U$.  If $U$ is the evolution
operator corresponding to some Hamiltonian, its eigenstates
are just those of the Hamiltonian.  This procedure will not
tell us what those eigenstates are, but we can perform 
measurements on the qudits in those states in order to gain 
information about them.  We may also simply be interested in
performing further operations on these states without measuring
them first, and we now have a way of producing them.

If singlet states are combined with the phase-estimation
algorithm for finding eigenvalues of $U$, we can, in a
certain sense, diagonalize the operator.  We saw that for
qubits we both knew the eigenvalues, at least to a level of
approximation that we can determine, and we produced qubits
in states that are very close to the eigenstates of $U$.  This
procedure should generalize to qudits.

\section*{Acknowledgments}
This research was supported by the National Science Foundation
under grant PHY-9970507, and by the European Union project
EQUIP under contract IST-1999-11053.

\bibliographystyle{unsrt}

\begin{thebibliography}{99}
\bibitem{kitaev} A.\ Kitaev, quant-ph/9511026.
\bibitem{cleve} R.\ Cleve, A.\ Ekert, C.\ Macchiavelo, and 
M.\ Mosca, Proc.\ R.\ Soc. Lond.\ A {\bf 454}, 339 (1998)
and quant-ph/9708016.
\bibitem{abrams}D.\ S.\ Abrams and S.\ Lloyd, Phys.\ Rev.\ Lett.\ 
{\bf 83}, 5162 (1999).
\bibitem{milburn}B.\ C.\ Travaglione and G.\ J.\ Milburn, 
quant-ph/0008053.
\bibitem{ivanovic}  I.\ D.\ Ivanovic, Phys.\ Lett. A {\bf 123}, 
257 (1987).
\bibitem{dieks} D.\ Dieks, Phys.\ Lett. A {\bf 126} 303 (1988).
\bibitem{peres} A.\ Peres, Phys.\ Lett. A {\bf 128}, 19 (1988).
\bibitem{georgi}See, for example, H.\ Georgi, \emph{Lie Algebras
in Particle Physics: from Isospin to Unified Theories} (Benjamin,
Reading, 1982), p. 114.
\end{thebibliography}

\end{document}